# Nonlocal Correlation in Heisenberg Spin Models

Muthuganesan R and R. Sankaranarayanan

*Department of Physics, National Institute of Technology, Tiruchirappalli– 620015, TamilNadu, India.*

**Abstract:** In this paper we investigate the nonlocal correlation (beyond entanglement) captured by measurement induced nonlocality and geometric quantum discord for a pair of interacting spin-1/2 particles at thermal equilibrium. It is shown that both the measures are identical in measuring the correlation. We show the possibility of nonlocal correlation between the spins without entanglement, which ceases to exist only for the maximal mixture of product bases. We also observe that while interaction between the spins is responsible for the enhancement of correlation, this non-classicality decreases with the intervention of external magnetic field.

## 1. Introduction

The results of local measurements on some composite quantum systems cannot be reproduced classically. This is known as quantum nonlocality, which is a valuable resource for information processing and has no analogy in the classical world. Nonlocality of any quantum state can be demonstrated quantitatively through the violation of Bell inequality[1]. Entanglement, exhibits in composite quantum system as a consequence of the superposition principle, is originally introduced by Einstein and his co-workers[2] and Schrödinger[3,4]. Since the work of Bell, entanglement is believed to be the manifestation of nonlocality which cannot be accounted by any local hidden variable model. Entanglement is also widely known as the basis for various information processing task[5-9]. Though all pure entangled states are nonlocal, in terms of violation of Bell inequality, all mixed entangled states are not. In fact, Werner showed through an explicit example that mixed entangled state can also obey Bell or Bell like inequalities[10]. This is attributed to the presence of noise or mixedness, which are responsible in destroying nonlocal correlation between different parts of the composite system, and hence some of the mixed entangled states behave locally[11].

Entanglement may not be the ultimate resource behind the power of quantum computation[12]. It has been recently found that, even without entanglement some mixed-state based schemes such as the so-called deterministic quantum computation with one quantum bit (DQC1)[13] allow to improve the performance of computing tasks[14]. More generally, separable states possess genuine quantum correlation[15] captured by quantum discord[16,17]. Quantum discord is the first measure to quantify the correlations that are beyond the entanglement. Two classically identical expressions for the mutual information between two parties generally differ in the quantum regime. This difference defines the *quantum discord*. In view of this, it is now broadly accepted that entanglement is *not* the complete manifestation of nonlocality. However, the calculation of quantum discord involves complex optimization procedure in evaluating the underlying classical correlations[18].

In order to overcome this difficulty, a geometric measure of quantum discord (GMOD) has been introduced[19], which measures the distance between a given state and the set of zero-discord states. Evaluation of this measure does not require any optimization, and an analytical expression of GMOD for an arbitrary state is also available[20]. Further, S. Luo and S. Fu defined the maximum nonlocal effect of locally invariant measurements as measurement induced nonlocality (MIN)[21]. This measure is in some sense dual to the GMOD[19], and quantifies the global effect of statistical mixture of states rather than the effect due to entanglement. Hence the above notion of nonlocality may also be considered to be more general than the Bell's version of nonlocality. This quantity has been investigated for bound entangled states[22], general bipartite system[23] and anisotropic Heisenberg spin chain[24]. An entropic version of MIN, the maximal increment of von Neumann entropy due to locally invariant measurements, has also been studied elsewhere[25]. Recently, the quantity MIN is also proposed based on trace distance[26] and skew information[27].

In this paper, we study the behaviour of nonlocal correlation for thermal state of a pair of interacting spin-1/2 particles using MIN and GMOD along with the entanglement. In particular, we consider the isotropic spins with DM interaction[28, 29] and XXZ model in presence of external homogeneous magnetic field. It is shown that MIN is twice of GMOD for the above system, implying that both the quantities are identical in measuring nonlocal correlations. It is further shown that MIN (or GMOD) is a useful measure of quantum correlation even if the spins are unentangled. We also observed that while nonlocal correlation is enhanced with the interaction between the spins at low temperature, presence of external field tends to supress the correlation as the system becomes more classical.

## 2. Preliminaries

### 2.1 Entanglement

The amount of entanglement associated with a given two-qubit state $\rho$ can be quantified using concurrence[30], which is defined as $C(\rho) = \max\{0, \lambda_1 - \lambda_2 - \lambda_3 - \lambda_4\}$ where $\lambda_i$ are square root of eigenvalues of matrix $R = \rho\tilde{\rho}$ arranged in decreasing order. Here $\tilde{\rho}$ is spin flipped density matrix, which is defined as $\tilde{\rho} = (\sigma_y \otimes \sigma_y)\rho^*(\sigma_y \otimes \sigma_y)$. The symbol * denotes complex conjugate in computational basis. It is known that $0 \leq C(\rho) \leq 1$ with minimum and maximum values correspond to the separable and maximally entangled states respectively.

### 2.2 Geometric Measure of Quantum Discord

An arbitrary bipartite state $\rho$ in $\mathbb{C}^2 \otimes \mathbb{C}^2$ can be represented as,

$$\rho = \frac{\mathbb{I}^A \otimes \mathbb{I}^B}{4} + \sum_{i=1}^{3} x_i X_i \otimes \frac{\mathbb{I}^B}{\sqrt{2}} + \sum_{j=1}^{3} \frac{\mathbb{I}^A}{\sqrt{2}} \otimes y_j Y_j + \sum_{i,j} t_{ij} X_i \otimes Y_j \qquad (1)$$

where the matrices $X_i$ and $Y_j$ are orthonormal Hermitian operator bases associated with the subsystems $A$ and $B$ respectively such that $tr(X_k X_l) = tr(Y_k Y_l) = \delta_{kl}$. The components of Bloch vectors are $x_i = tr[\rho(\sigma_i \otimes \mathbb{I}^B)]/2$ and $y_j = tr[\rho(\mathbb{I}^A \otimes \sigma_j)]/2$ with $t_{ij} = tr[\rho(\sigma_i \otimes \sigma_j)]/2$ being real matrix elements of correlation matrix $T$.

The geometric measure of quantum discord of state $\rho$ quantifies the nonlocal correlation through the least Hilbert-Schmidt distance between the given state and zero discord state. Mathematically it is defined as[19],

$$D(\rho) = \min_{\chi \in \Omega} \|\rho - \chi\|^2 \qquad (2)$$

where $\Omega$ is a set of all zero discord states, $\|A\|^2 = tr(A^\dagger A)$ is the square of Hilbert-Schmidt norm of an operator $A$ and $\chi = \sum_{i=1}^{2} p_i |\psi_i\rangle\langle\psi_i| \otimes \rho_i$ is a zero discord state on Hilbert space $H^A \otimes H^B$ with probability distribution $\{p_i\}$. It has a closed formula for any general bipartite system as given by[20]

$$D(\rho) = 2(trS - \max(k_i))$$

where the matrix $S = \frac{1}{4}(xx^t + TT^t)$, with $x^t = (x_1 \; x_2 \; x_3)$ and $k_i$ are the eigenvalues of $S$ and superscript $t$ denotes the transpose of a matrix. After the optimization $D(\rho)$ has the following very tight lower bound

$$Q(\rho) = \frac{2}{3}\left(2trS - \sqrt{6tr(S^2) - 2(trS)^2}\right). \quad (3)$$

This quantity satisfies all the criteria for being a meaningful measure of quantum correlation. Thus $Q(\rho)$ is also experimentally convenient, since one does not need to perform a full tomography of the state.

### 2.3 Measurement Induced Nonlocality

Another measure of nonlocality (or quantum correlation) in a given system, as introduced by S. Fu and S. Luo, is a measurement based quantity which is dual to $D(\rho)$ called measurement induced nonlocality (MIN). It is defined as the maximum of square of Hilbert–Schmidt distance between pre- and post-measurement state[21]:

$$N(\rho) = \max_{\Pi^A} \|\rho - \Pi^A(\rho)\|^2 \quad (4)$$

where the maximum is taken over all local projective measurements. Here $\Pi^A(\rho) = \sum_k (\Pi_k^A \otimes \mathbb{I}^B) \rho (\Pi_k^A \otimes \mathbb{I}^B)$, with $\{\Pi_k^A\} = \{|k\rangle\langle k|\}$ being the projective measurements on the subsystem $A$, which do not change $\rho^A$ locally (i.e., $\sum_k \Pi_k^A \rho^A \Pi_k^A = \rho^A$). For a general bipartite state, MIN is evaluated for $\mathbb{C}^2 \otimes \mathbb{C}^d$ dimensional systems and the closed formula is given by

$$N(\rho) = \begin{cases} tr(TT^t) - \frac{1}{\|x\|^2} x^t TT^t x & if\ x \neq 0 \\ tr(TT^t) - \lambda_{min} & if\ x = 0 \end{cases} \quad (5)$$

where $\lambda_{min}$ is the minimum eigenvalue of the matrix $TT^t$. It is known that MIN has the maximum value of 0.5 for the EPR (maximally pure entangled) states.

### 3. Heisenberg Isotropic Model

First, we consider a system of two spin- ½ particles (electrons) having Heisenberg isotropic and DM interactions, whose Hamiltonian is given by,

$$H = \frac{1}{2}\left[J \sum_i \sigma_1^i \sigma_2^i + \vec{D}.(\vec{\sigma_1} \times \vec{\sigma_2})\right] \quad (6)$$

where $\sigma_k^i$ ($k = 1,2\ and\ i = x,y,z$) are Pauli spin matrices, $J$ is the exchange coupling constant, and $\boldsymbol{D}$ is DM vector which we choose to be along the z-axis, (i.e., $\boldsymbol{D} = D\hat{\boldsymbol{z}}$)

Eigenvalues of the Hamiltonian are given by $E_{1,2} = J/2$ and $E_\pm = -J/2 \pm \eta$ with corresponding eigenvectors $|\varphi_1\rangle = |00\rangle$, $|\varphi_2\rangle = |11\rangle$ and $|\varphi_\pm\rangle = [((\pm\eta/(J - iD))|01\rangle + |10\rangle]/\sqrt{2}$ respectively with $\eta = \sqrt{J^2 + D^2}$. For $D = 0$, the Hamiltonian reduces to Heisenberg isotropic interaction spin model and the eigenfunctions $|\varphi_\pm\rangle$ are reduced to the maximally entangled EPR pairs.

The density matrix describing the system (6) at an equilibrium temperature $T$ is $\rho(T) = Z^{-1} \exp(-H/kT)$, where $Z = tr[\exp(-H/kT)]$ is the partition function and $k$ is the Boltzmann's constant. In the computational basis the density matrix takes the form

$$\rho(T) = \frac{1}{Z}\begin{pmatrix} \mu & 0 & 0 & 0 \\ 0 & \omega & v & 0 \\ 0 & v^* & \omega & 0 \\ 0 & 0 & 0 & \mu \end{pmatrix} \quad (7)$$

where $\mu = e^{-J/2kT}$, $\omega = e^{J/2kT}\cosh(\eta/kT)$, $v = -(J+iD)e^{J/2kT}\sinh(\eta/kT)/\eta$ and $Z = 2e^{-J/2kT}[e^{J/kT}\cosh(\eta/kT) + 1]$. The concurrence, MIN and GMOD for the above state are computed as

$$C(\rho) = \frac{2}{Z}\max\{0, |v| - \mu\} \quad (8.a)$$

$$N(\rho) = \frac{2}{Z^2}|v|^2. \quad (8.b)$$

$$Q(\rho) = \frac{1}{Z^2}|v|^2 = \frac{1}{2}N(\rho). \quad (8.c)$$

It is interesting to note that MIN and GMOD are proportional to each other for the above Hamiltonian. In what follows, we compare both entanglement and nonlocal correlation (beyond entanglement) for the thermal state given by eq. (7) as a function of $J$. To study the influence of exchange interaction, we set the DM interaction to be zero ($D = 0$). From Fig. 1(a), we observe that $C(\rho) = 0$ for $J \leq J_c = kT \ln 3/2$. On the other hand, the measures MIN and GMOD are non-zero except at $J = 0$, implying that there exists quantum correlation between the spins even if they are not entangled. We also observe that while the spins are weakly correlated in ferromagnetic phase ($J < 0$) than that in the antiferromagnetic phase ($J > 0$). This could be a possible indication of change of phase of the magnetic system. A similar observation is made in such spin systems in terms of quantum discord as well[31]. At $J = 0$, $\rho = \mathbb{I}_4/4$ (maximally mixed state) for which $N(\rho) = Q(\rho) = 0$ implies that nonlocal correlation is essentially induced by the exchange interaction. For $J > J_c$, the concurrence increases to the maximum when $J \gtrsim 3kT$. In other words, entanglement is maximum in the antiferromagnetic phase provided the exchange interaction strength between the spins is larger than $3kT$. Fig. 1(a) also shows the region of maximum entanglement where quantum correlation between the spins is maximum.

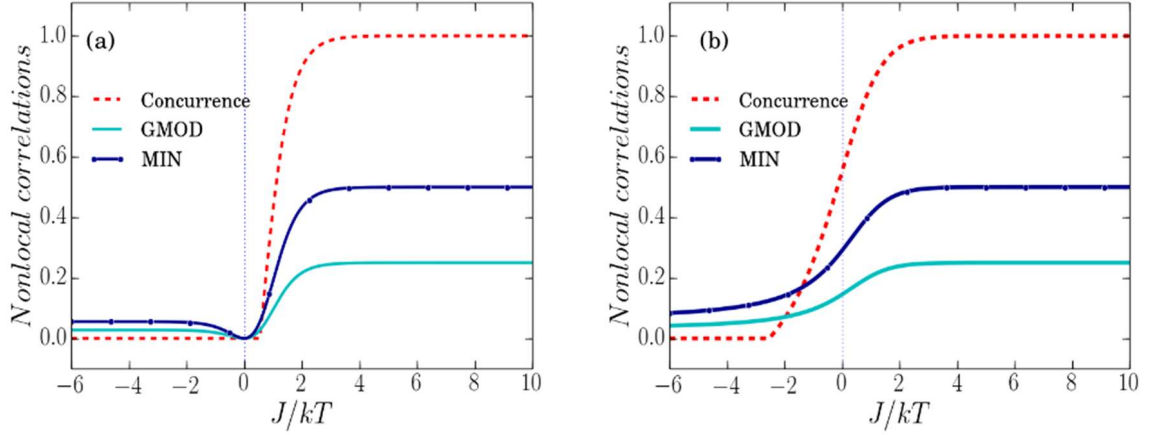

**Figure 1**: The nonlocal correlation in Heisenberg isotropic interaction for the Hamiltonian (6) with (a) $D/kT = 0$ and (b) $D/kT = 2$.

In order to see the effect of DM interaction on correlation, we have taken $D = 2kT$. It is clearly observed from Fig. 1(b) that $C(\rho) = 0$ for $J \lesssim -2.55\,kT$. For $J > -2.55\,kT$, the concurrence gradually increases to the maximum when $J \gtrsim 3kT$. On comparing this with Fig. 1(a), we also observe that entanglement is enhanced by the DM coupling, resulting to the enhancement of MIN and GMOD when $|J/kT|$ is small. Here also maximum entanglement resulting to maximum correlation when $J \gtrsim 3kT$.

## 4. Heisenberg anisotropic Model

Here we consider two electron system having anisotropic Heisenberg interaction in presence of an external homogeneous magnetic field. The Hamiltonian of the model is given by

$$H = \frac{1}{2}\left[J\left(\sigma_1^x \sigma_2^x + \sigma_1^y \sigma_2^y + (1+\Delta)\sigma_1^z \sigma_2^z\right) + B(\sigma_1^z + \sigma_2^z)\right] \quad (9)$$

where $\Delta$ is the dimensionless anisotropy parameter along z-axis and $B$ denotes the magnetic field along $z$ direction. Eigenvalue of the Hamiltonian are given by $E_{1,2} = J(1+\Delta)/2 \pm B$, and $E_{3,4} = -J(1+\Delta)/2 \pm J$ with corresponding eigenvectors $|\psi_1\rangle = |00\rangle$, $|\psi_2\rangle = |11\rangle$ and $|\psi_{3,4}\rangle = (|01\rangle \pm |10\rangle)/\sqrt{2}$ respectively. The corresponding thermal state is given by,

$$\rho(T) = \frac{1}{Z}\begin{pmatrix} \delta_+ & 0 & 0 & 0 \\ 0 & \epsilon & \kappa & 0 \\ 0 & \kappa & \epsilon & 0 \\ 0 & 0 & 0 & \delta_- \end{pmatrix} \quad (10)$$

where $\delta_\pm = e^{-(\alpha \pm B/kT)}$, $\epsilon = e^\alpha \cosh(J/kT)$, $\kappa = -e^\alpha \sinh(J/kT)$ and $Z = e^{-\alpha}\cosh(B/kT) + e^\alpha \cosh(J/kT)$ with $\alpha = J(1+\Delta)/2kT$. The concurrence, MIN and GMOD of the state are computed as

$$C(\rho) = \frac{2}{Z}\max\left\{0, |\kappa| - \sqrt{\delta_+ \delta_-}\right\} \tag{11.a}$$

$$N(\rho) = \frac{2}{Z^2}\kappa^2. \tag{11.b}$$

$$Q(\rho) = \frac{1}{Z^2}\kappa^2 = \frac{1}{2}N(\rho) \tag{11.c}$$

showing again the simple linear relation between MIN and GMOD.

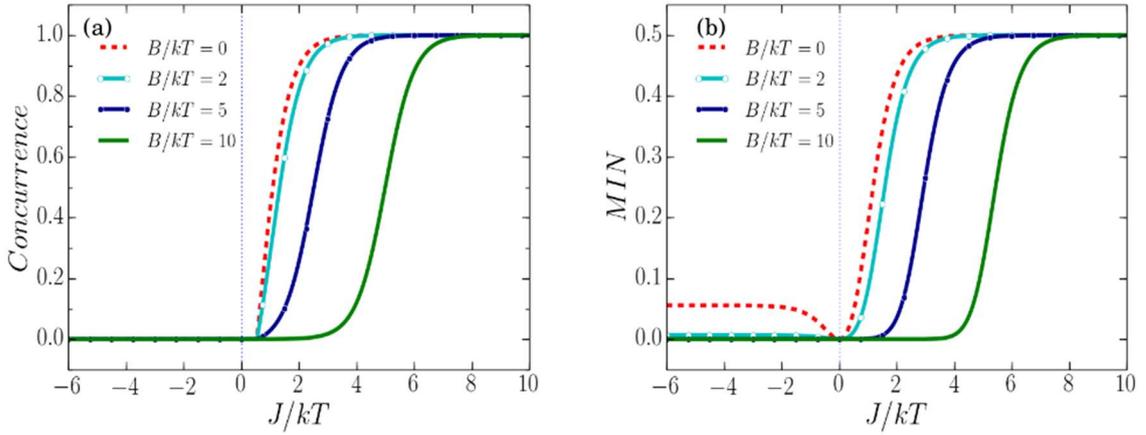

Figure 2: (a) Concurrence and (b) MIN for isotropic Heisenberg spins system with various magnetic field.

To investigate the entanglement and quantum correlation of the above system, first we consider the isotropic interaction by setting $\Delta = 0$. As shown in Fig. 2(a), in presence of the external field ($B \neq 0$) the spins remain unentangled for $J \leq J_c$. On the other hand, for $J > J_c$ as the field increases the entanglement decreases. Similarly, it is seen from Fig. 2(b) that the correlation measure MIN (and hence GMOD) is decreasing with the increase of field. Hence the magnetic field is always resulting to decrease the entanglement as well as the quantum correlation between the spins. Here also we observe that maximum entanglement corresponds to the maximum correlation between the spins. From the equation (11), it also clear that reversing the direction of $B$ will leave the entanglement, MIN and GMOD of the system unaffected.

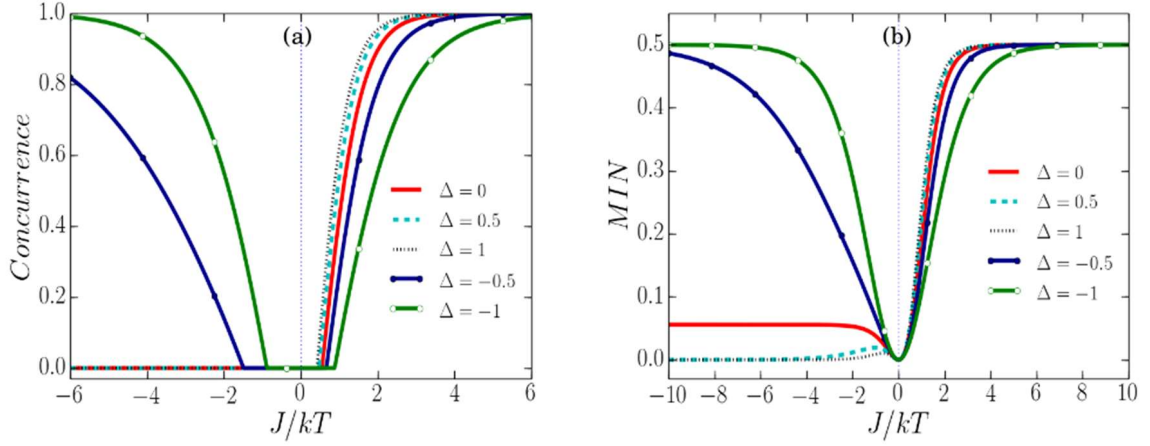

Figure 3: (a) Concurrence and (b) MIN for the Hamiltonian (9) without magnetic filed.

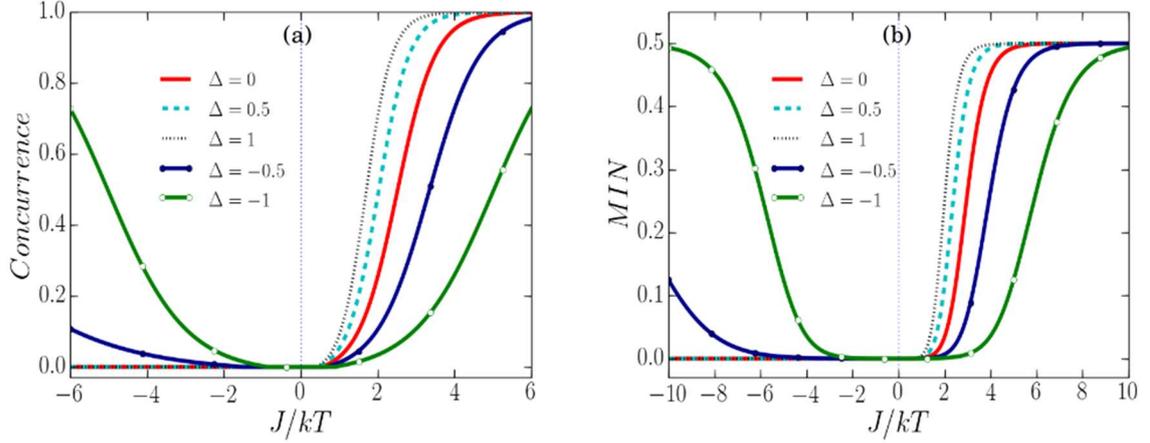

Figure 4: (a) Concurrence and (b) MIN for the Hamiltonian (9) with the magnetic field $B/kT = 5$.

Now we look at the role of anisotropic parameter Δ on entanglement and MIN at zero field. It is clear from Fig. 3(a) that while entanglement of the spins for Δ > 0 is very similar to that of Δ = 0, entanglement for Δ < 0 is significantly different. In particular, the entanglement is induced for the latter case in the ferromagnetic phase. In other words, the entanglement is largely induced in the ferromagnetic phase if the spin interactions are stronger in the *xy*-plane. In Fig. 3(b) we show the behaviour of MIN for different Δ. We notice here that unlike concurrence, MIN is more sensitive with the anisotropic parameter particularly in the ferromagnetic phase. The nonlocal correlation between the spins are also found to be maximum in the limit $|J|/kT \to \infty$ i.e., $T \to 0$ for finite $J$. In presence of external field ($B \neq 0$) the concurrence and MIN are shown to be decreasing as shown in Fig. 4. The decrease in nonlocal correlation with the increase of field implying that the system loses the quantum signature and becomes more classical.

5. Conclusions

In this paper we have studied the behaviour of entanglement measured by concurrence along with nonlocal correlation captured by measurement induced nonlocality (MIN) and geometric measure of quantum discord (GMOD) for a pair of interacting spin-1/2 particles at thermal equilibrium. Since we observe that MIN is just twice of GMOD for the system under investigation, it is enough to look into the MIN as the measure of correlation. In the absence of external magnetic field, unlike entanglement the MIN is found to be zero only when the state is maximally mixed, implying that correlation is induced essentially by the Heisenberg exchange interaction of the spins. In other words, though quantum correlation between the spins exists in the absence of entanglement, the correlation is found to be maximum when the spins are maximally entangled. It is also observed that though the anisotropic interaction between the spins induce both entanglement and MIN, the latter quantity is more sensitive with the anisotropic parameter. The intervention of external magnetic field decreases entanglement and MIN such that the correlation without entanglement is negligible. These results indicate that MIN and GMOD are useful measure of quantum correlation (non-classicality) in the absence of entanglement.